\documentclass[lettersize,journal]{IEEEtran}
\usepackage{amsmath,amsfonts}
\usepackage[ruled, vlined, linesnumbered]{algorithm2e}
\usepackage{float}
\usepackage{array}
\usepackage[caption=false,font=normalsize,labelfont=sf,textfont=sf]{subfig}
\usepackage{textcomp}
\usepackage{stfloats}
\usepackage{url}
\usepackage{verbatim}
\usepackage{graphicx}
\usepackage{cite}
\usepackage{amsthm}
\usepackage{booktabs}%
\usepackage{multirow}%
\usepackage{enumerate}%
\usepackage{stfloats}
\usepackage{amssymb}
\usepackage{xcolor}
\graphicspath{{./Figures/}}% Images path
\newtheorem{definition}{Definition}
\newtheorem{remark}{Remark}
\newtheorem{lemma}{Lemma}
\newtheorem{theorem}{Theorem}

\begin{document}

\title{TDC-Cache: A Trustworthy Decentralized Cooperative Caching Framework for Web3.0}

\author{
	Jinyu~Chen,
	Long~Shi,
	Taotao~Wang,
	Jiaheng~Wang,
	and Wei~Zhang
% <-this % stops a space
\thanks{Jinyu Chen, Long Shi are with the School of Electronic and Optical Engineering, Nanjing University of Science and Technology, Nanjing 210094, China (e-mail: Jinyu.Chen@njust.edu.cn, slong1007@gmail.com). Taotao Wang is with the College of Electronics and Information Engineering, Shenzhen University, Shenzhen 518052, China (e-mail: ttwang@szu.edu.cn). Jiaheng Wang is with the National Mobile Communications Research Laboratory, Southeast University, Nanjing 211102, China, and also with the Purple Mountain Laboratories, Nanjing 211111, China (e-mail: jhwang@seu.edu.cn). Wei Zhang is with the School of Electrical Engineering and Telecommunications, The University of New South Wales, Sydney, NSW 2052, Australia (e-mail: w.zhang@unsw.edu.au).}
}



\maketitle

\begin{abstract}
	The rapid growth of Web3.0 is transforming the Internet from a centralized structure to decentralized, which empowers users with unprecedented self-sovereignty over their own data.
	However, in the context of decentralized data access within Web3.0, it is imperative to cope with efficiency concerns caused by the replication of redundant data, as well as security vulnerabilities caused by data inconsistency.
	To address these challenges, we develop a Trustworthy Decentralized Cooperative Caching (TDC-Cache) framework for Web3.0 to ensure efficient caching and enhance system resilience against adversarial threats.
	This framework features a two-layer architecture, wherein the Decentralized Oracle Network (DON) layer serves as a trusted intermediary platform for decentralized caching, bridging the contents from decentralized storage and the content requests from users.
	In light of the complexity of Web3.0 network topologies and data flows, we propose a Deep Reinforcement Learning-Based Decentralized Caching (DRL-DC) for TDC-Cache to dynamically optimize caching strategies of distributed oracles.
	Furthermore, we develop a Proof of Cooperative Learning (PoCL) consensus to maintain the consistency of decentralized caching decisions within DON.
	Experimental results show that, compared with existing approaches, the proposed framework reduces average access latency by 20\%, increases the cache hit rate by at most 18\%, and improves the average success consensus rate by 10\%.
	Overall, this paper serves as a first foray into the investigation of decentralized caching framework and strategy for Web3.0.
\end{abstract}


\section{Introduction}
\IEEEPARstart{T}{he} rapid evolution of Web3.0 is driving the transition from traditional centralized systems to decentralized architectures.
Leveraging blockchain, decentralized storage, and smart contracts, Web3.0 empowers users with unprecedented self-sovereignty over their own data through Decentralized Applications (DApps) \cite{kim2023moving}.
As the variety of DApps boosts, user demands for distributed data access with low latency, high throughput, and high scalability are growing significantly.
Nevertheless, current decentralized storage struggles to fulfill these demands due to inherent limitations within decentralized architectures.

Decentralized storage is significantly constrained by critical inefficiencies in terms of high network latency stemmed from distributed consensus and inefficient content retrieval caused by chunked data storage \cite{benisi2020blockchain}.
These inefficiencies inevitably introduce significant latency in distributed data access, significantly degrade the responsiveness of DApps, and negatively impact user Quality of Service \cite{jiang2018low}.
In contrast, Web2.0 addresses similar challenges through hierarchical caching frameworks (e.g., Content Delivery Network) that store frequently accessed data closer to users, significantly reducing the latency of data retrieval and improving the responsiveness of user requests \cite{smith1982cache}.
Commonly, these frameworks have developed diverse caching strategies to explore skewed distributions of content popularity, where a small fraction of content accounts for the majority of access requests \cite{breslau1999web}.

Traditional caching strategies, such as Least Recently Used (LRU), First-In-First-Out (FIFO), and Least Frequently Used (LFU), prioritize simplicity but fail to fully utilize this skewness.
To address these limitations, advanced strategies such as Leave Copy Down (LCD) \cite{laoutaris2006lcd} and TinyLFU \cite{einziger2017tinylfu} have emerged.
Specifically, LCD improves LRU by placing content close to users to reduce retrieval distance.
Similarly, TinyLFU optimizes LFU by using approximate frequency statistics to improve space utilization.
Particularly, these strategies improve efficiency but remain centralized in design.
In contrast, decentralized architecture of Web3.0, characterized by distributed data ownership and dynamic content interactions, demands adaptive caching strategies that support real-time optimization through decentralized decision-making mechanisms \cite{cao2025web}.
Inspired by the hierarchical caching frameworks of Web2.0, a potential solution is to employ an intermediary caching layer bridging users and decentralized storage.
Due to fundamental differences between network architectures, centralized caching strategies for Web2.0 cannot be directly applied to the decentralized architecture of Web3.0.
It is necessary to develop entirely new caching solutions for Web3.0 from scratch.
However, related research in Web3.0 remains underexplored and warrants further investigation.

In recent years, Deep Reinforcement Learning (DRL) has emerged to optimize caching decisions in dynamic environments.
By evaluating dynamic factors such as user behavior, content popularity, and network conditions, DRL-based caching strategies dynamically update cached content to prioritize frequently requested data.
Leveraging real-time feedback of cache hits or misses, these strategies are continuously refined to adapt to dynamic environments and thereby ensure long-term caching efficiency.
Existing works of \cite{wu2019dynamic,li2021deep,wang2024towards} have demonstrated that DRL-based caching strategies outperform traditional static methods by significantly improving cache hit rates and reducing content retrieval latency.

In this context, DRL-based caching strategies hold promise for the caching of Web3.0, as its decentralized architecture provides computational resources for training DRL models \cite{zhang2019joint} and storage capacity for caching popular content \cite{fan2022dlbn}.
However, several challenges hinder the applications of these strategies in Web3.0.
First, decentralized infrastructure of Web3.0 poses challenges for DRL training and inference, which traditionally depend on centralized or semi-centralized coordination \cite{shen2024artificial, niu2025blockchain}.
% This limitation hinders the large-scale implementation of DRL across decentralized networks.
Second, the scalability of distributed training in Web3.0 is impeded by node heterogeneity in limited computational resources and bandwidth constraints \cite{zhu2025moe, verbraeken2020survey}.
Notably, the training process of DRL models rely on extensive data sharing (e.g., state transitions, actions, and rewards) and a global view for policy optimization, both of which are challenging to be achieved in decentralized environments.
Third, Byzantine nodes pose significant security risks by tampering with data, gradients, or model parameters during model training \cite{guerraoui2024byzantine}.
In the course of our literature survey, it has been identified that the existing Web3.0 frameworks are deficient in providing robust solutions against the aforementioned threats.

Driven by these challenges, we propose a two-layer framework called Trustworthy Decentralized Cooperative Caching (TDC-Cache) to enhance the efficiency and security of distributed data access in Web3.0.
The novelty of the proposed framework lies in introducing a trusted intermediary between users and decentralized storage to accelerate data retrieval.
Building upon this framework, we develop a DRL-Based Decentralized Caching (DRL-DC) strategy to optimize caching efficiency and a Proof of Cooperative Learning (PoCL) consensus to ensure the consistency of distributed data against adversarial threats.
Our primary contributions are as follows:

\begin{itemize}
	\item We design the TDC-Cache framework for Web3.0.
	In this framework, the Decentralized Oracle Network (DON) layer serves as a trusted intermediary platform for decentralized caching, bridging the contents from decentralized storage and the content requests from users.
	Extending beyond the current state of the art, the proposed framework represents the pioneering effort that incorporates caching mechanisms within the Web3.0 architecture to augment the efficiency and security of data retrieval.

	\item We propose the DRL-DC strategy to dynamically optimize caching decisions across DON.
	To address uncertainty and partial observability of distributed user requests and caches, we formulate the caching problem as a Decentralized Partially Observable Markov Decision Process and solve this problem by a QMIX-based algorithm.
	This algorithm maximizes long-term cache hit rates and enhances the efficiency of data retrieval through decentralized oracle cooperation under dynamic content requests of users and limited storage capacity of oracles.

	\item We propose the PoCL consensus for DON to defend against malicious oracles that falsify cached data and disrupt the consistency of decentralized caching strategy.
	This consensus requires oracles in DON to submit verifiable cooperative proofs, from which a training oracle is selected to aggregate the proofs and subsequently train the caching strategy.
	We theoretically evaluate the safety, liveness, and success consensus rate of PoCL. Analytical results demonstrate that PoCL maintains the consistency of decentralized caching strategy and enhances the security of decentralized learning.
	
	\item Experimental results demonstrate that the proposed framework reduces average access latency by 20\% compared with direct access. Furthermore, the DRL-DC strategy improves the cache hit rate by at most 18\% compared with traditional strategies, and the PoCL consensus achieves a higher success rate of 10\% than Practical Byzantine Fault Tolerance (PBFT).
\end{itemize}

The remainder of this paper is structured as follows: Section \ref{sec:Related Work} reviews related works, Section \ref{sec:Proposed TDC-Cache Framework} describes the proposed framework, Section \ref{sec:System Model} presents the system model, 
Section \ref{sec:Latency Optimization of TDC-Cache} studies the latency optimization problem, 
Section \ref{sec:Proof of Cooperative Learning} details the design of PoCL, Section \ref{sec:Performance Analysis of Consensus} analyzes the performance of PoCL, Section \ref{sec:Experimental Results} presents experimental results, and Section \ref{sec:Conclusion} concludes this paper.
For convenience, key notations are summarized in Table \ref{tab:notice}.

\begin{table}[!t]
  \caption{Summary of Notations}
  \label{tab:notice}
	\centering
	\begin{tabular}{l l}
	\hline
	Notations & Definitions \\
	\hline
	$\mathcal{T}$ & Set of time slots \\
	$\mathcal{F}$ & Set of contents \\
	$\mathcal{M}$ & Set of oracles \\
	$\mathcal{U}$ & Set of CRs \\

	$t$ & Time slot index in $\mathcal{T}$\\
	$f$ & $f$-th content in $\mathcal{F}$ \\
	$m$ & $m$-th oracle in $\mathcal{M}$ \\
	$u$ & $u$-th content requester in $\mathcal{U}$ \\

	$C_{m}$ & Capacity of oracle $m$ \\

	$G_{m, u}^t$ & Content request sequence of $u$ to $m$ at time slot $t$ \\
	$g_{m, u, k}^t$ & $k$-th request of $G_{m, u}^t$ \\

	$O_{m}^t$ & Observation of oracle $m$ at time slot $t$ \\
	$A_{m}^t$ & Decision matrix of oracle $m$ at time slot $t$ \\
	$\tau_{m}^t$ & Action-observation history of oracle $m$ \\

	$\boldsymbol{S}^t$ & Joint states at time slot $t$\\
	$\boldsymbol{A}^t$ & Joint actions at time slot $t$\\
	$\boldsymbol{\tau}^t$ & Joint action-observation history at time slot $t$\\

	% $D_{g_{m, u, k}}^t$ & Retrieval latency for $g_{m, u, k}^t$ from oracles network \\
	% $D_{\text{ds}, g_{m, u, k}}^t$ & Retrieval latency for $g_{m, u, k}^t$ from decentralized storage\\
	% $D^t$ & Overall content retrieval latency at time $t$ \\

	% $\lvert \cdot \rvert$ & The number of elements in $\cdot$ \\
	\hline
	\end{tabular}
\end{table}

\section{Related Work}
\label{sec:Related Work}

In this section, we review the literature on different caching scenarios and consensus in decentralized networks.

\subsection{Caching Strategies in Different Domains}

Caching strategies are widely employed across different domains to reduce data access latency and mitigate network congestion.
For instance, \cite{wang2021cooperative} proposes a three-stage cooperative caching framework with dynamic content request prediction to optimize edge resource utilization in Internet of Vehicles (IoV).
This framework incorporates $k$-means clustering for stable vehicle group formation based on mobility patterns, employs Long Short-Term Memory networks for spatiotemporal content popularity prediction, and utilizes a DRL optimizer for real-time cache resource allocation under storage constraints.
As a follow-up work, \cite{zhou2023distributed} designs a distributed multi-agent DRL framework to minimize long-term content access costs and mitigate backhaul traffic in IoV.
This framework extends single-agent DRL to a multi-agent stochastic game for cooperative caching and employs deep $Q$-networks with neural network-driven Nash equilibrium approximation.
Later, \cite{wu2024cooperative} develops a cooperative edge caching scheme that employs federated learning for privacy-preserving model updates and utilizes multi-agent deep deterministic policy gradient for cooperative content placement among small-cell base stations.
However, these approaches depend on centralized controllers, rendering them unsuitable for decentralized architecture of Web3.0.
Furthermore, these frameworks lack security mechanisms to maintain data consistency and thwart against adversarial manipulations.

\begin{figure*}[!ht]
	\centering
	\includegraphics[width=\linewidth]{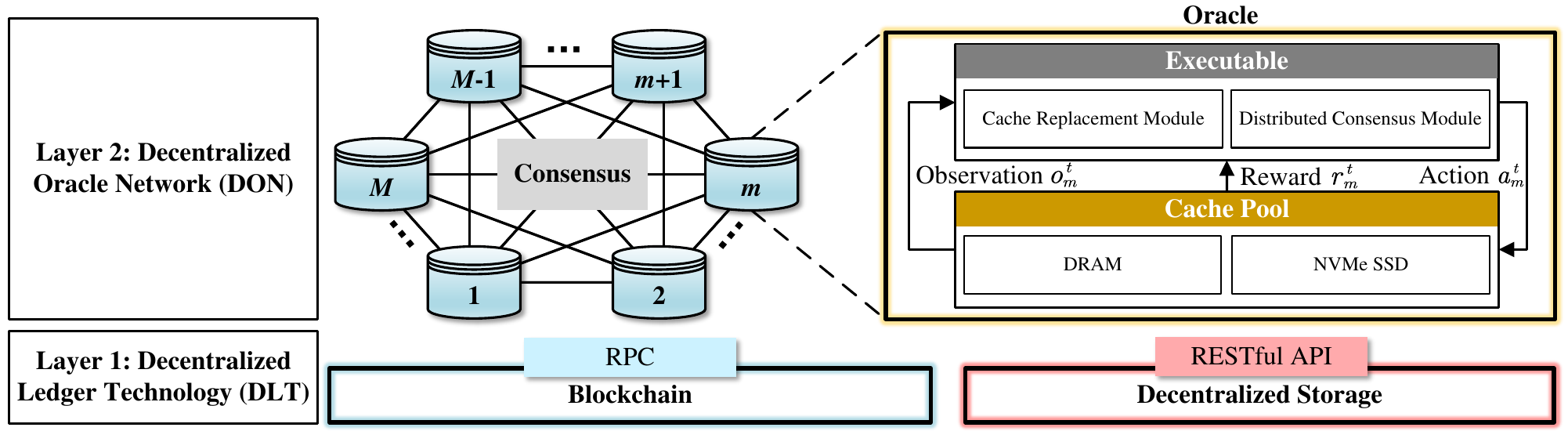}
	\caption{Architecture of TDC-Cache.}
	\label{fig:Architecture}
\end{figure*}

\subsection{Consensus in Decentralized Networks}

Distributed consensus are essential for maintaining data consistency and security in decentralized networks.
For instance, \cite{xu2023abc} develops a PBFT-based consortium blockchain to mitigate static network structures and high communication overhead.
This consensus ensures data consistency by employing artificial bee colony-optimized node preselection and preserves data integrity by replacing the commit phase of PBFT with optimized response cryptographic digest validation.
Moreover, \cite{qushtom2023two} proposes a two-stage PBFT architecture with trust-reward mechanisms for IoT to suppress malicious node behavior.
This consensus ensures data consistency by dynamically adjusting trust scores to penalize dishonest validators and leaders through reward reductions and deposit forfeitures.
Later, \cite{jiang2024proof} proposes a Proof-of-Trusted-Work consensus for decentralized IoT networks to tackle computing power centralization.
This consensus ensures data consistency through an on-chain reputation evaluation mechanism that verifies nodes’ computational contributions via embedded cryptographic proofs and dynamically adjusts block generation difficulty based on reputation thresholds to disincentivize mining pools.
In the context of DRL-aided decentralized caching, the design of distributed consensus to ensure the consistency of decentralized caching decisions remains challenging.

\section{Proposed TDC-Cache Framework}
\label{sec:Proposed TDC-Cache Framework}

In this section, we develop a novel decentralized caching framework for Web3.0 and introduce its architecture and workflow.

\subsection{The Architecture of TDC-Cache}
\label{sec:The Architecture of TDC-Cache}

Consider a typical Web3.0 infrastructure scenario with a blockchain network, a decentralized storage network, and distributed data accessible to users.
Targeting this infrastructure scenario, Fig. \ref{fig:Architecture} shows the proposed Trustworthy Decentralized Cooperative Caching (called TDC-Cache for short) framework.
This framework features a two-layer architecture, i.e., the Decentralized Ledger Technology (DLT) layer and the Decentralized Oracle Network (DON) layer.
In this paper, we refer to the data as \textit{content}, the decentralized storage as \textit{content providers} (CP), and the users as \textit{content requesters} (CR), respectively.

\subsubsection{The DLT Layer}

This layer, comprising the blockchain and decentralized storage, provides a trustworthy and secure platform for TDC-Cache.

\begin{itemize}
	\item \textbf{Blockchain} builds up a secure execution environment for the Oracle Management Contract (OMC) and the Data Management Contract (DMC). Specifically, OMC manages oracle states (i.e., registration records and operational status), and DMC maintains the mapping between cached content and content identifiers (CID) generated by decentralized storage platforms.

	\item \textbf{Decentralized storage} (e.g., IPFS \cite{benet2014ipfs} and SWARM \cite{ethswarm}) stores large volumes of essential DApp data such as multimedia files and smart contract bytecode. To ensure interoperability between different decentralized storage platforms, DMC standardizes CIDs to promote the universality of the caching architecture.
	
\end{itemize}

\subsubsection{The DON Layer}

This layer comprises a group of distributed oracles that act as intermediaries between CR and CP.
Each oracle provides limited storage space for frequently accessed content and computational resources to optimize caching decisions.
An oracle consists of a cache pool and an executable component, wherein the executable component includes a caching module and a distributed consensus module.
The function of each module is outlined as follows:

\begin{itemize}
	\item \textbf{The caching module} manages content in the cache pool. This module downloads missing content from DLT and updates invalidated cache.
	\item \textbf{The distributed consensus module} implements the proposed consensus by coordinating data interactions among the oracles in DON.
	\item \textbf{The cache pool} comprises Non-Volatile Memory Express Solid State Drive (NVMe SSD) and Dynamic Random Access Memory (DRAM), wherein NVMe SSD stores cached data blocks (i.e., contents) of DApps, while DRAM holds hash-based indices associated with the cached content.
\end{itemize}

The DON layer communicates with DLT via Remote Procedure Call (RPC) and Representational State Transfer Application Programming Interface (RESTful API).
Specifically, the oracles in DON call smart contracts using RPC provided by the blockchain.
Additionally, contents are accessed through RESTful API provided by the decentralized storage.

\begin{remark}
	In the existing works \cite{deebak2023healthcare, zhu2024trustworthy, murray2023promise}, CRs only request content directly from DLT.
	However, the unstructured and decentralized storage mechanism of DLT makes content organization and retrieval inefficient.
	To address these issues, the proposed framework introduces DON, and allows CRs to access content through oracles in DON that learn dynamic data access pattern of Web3.0.
\end{remark}

\subsection{The Workflow of TDC-Cache}
\label{sec:Workflow}

\begin{figure*}[!t]
	\centering
	\includegraphics[width=\linewidth]{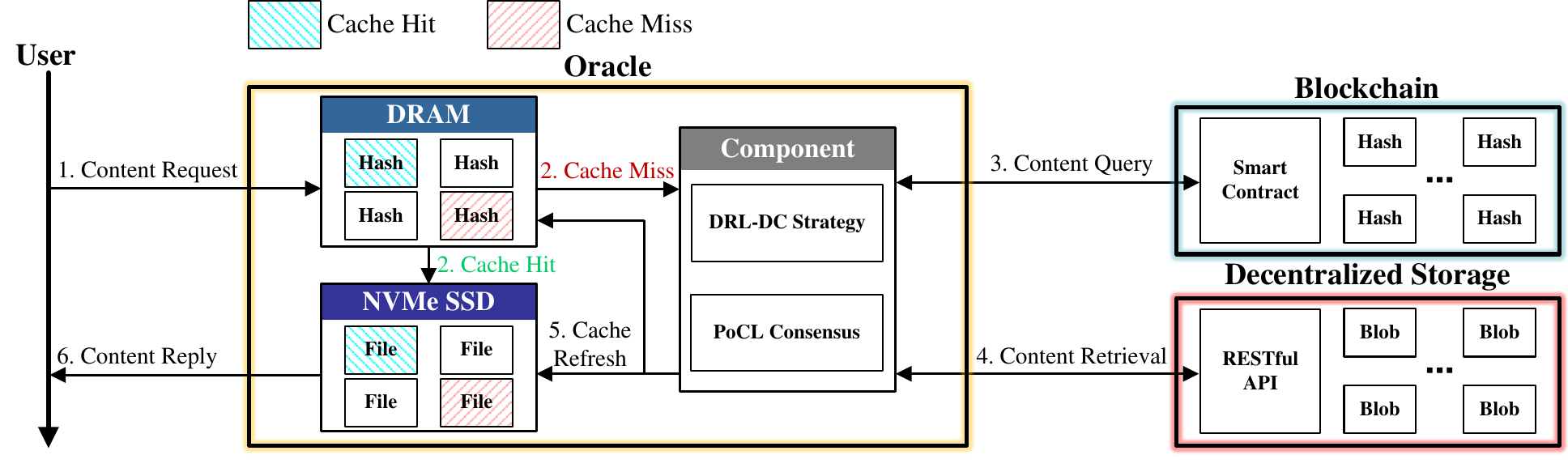}
	\caption{The workflow of TDC-Cache.}
	\label{fig:Workflow}
\end{figure*}

Fig. \ref{fig:Workflow} illustrates the caching workflow of the proposed framework.

\begin{enumerate}[Step 1: ]
	\item \textbf{content request}. When a DApp user generates a content request, the application forwards the request to the oracle.
	\item \textbf{request process}. Upon receiving the request, the oracle checks its cache pool for the corresponding content hash. If a cache hit occurs, proceed to Step 6; otherwise, proceed to Step 3.
	\item \textbf{content query}. The caching module queries the CID corresponding to the content hash by calling the DMC.
	\item \textbf{content retrieval}. The caching module downloads the content locally from the decentralized storage using the RESTful API.
	\item \textbf{cache refresh}. After downloading, the caching module updates the cache pool.
	\item \textbf{content reply}. Finally, the oracle delivers the requested content to the user.
\end{enumerate}

\section{System Model}
\label{sec:System Model}

Let us consider $M$ independent oracles, $F$ different contents, and $U$ CRs described in Section \ref{sec:The Architecture of TDC-Cache}.

\subsection{Content Popularity}
\label{sec:Content Popularity}

Let $\mathcal{F} = \{1, \ldots, f, \ldots, F\}$ be the set of $F$ contents, where $f$ is the $f$-th content on the DLT layer.
Each content has a finite data size.
In addition, the popularity of each content follows a Zipf-like distribution, i.e., $X \sim Zipf(\alpha, F)$, where $\alpha > 0$ is the power exponent that controls the distribution's steepness \cite{breslau1999web}.
The increase of $\alpha$ implies a more skewed content popularity distribution.
Given this distribution, the probability of the $k$-th ranked content being requested is given by
\begin{equation}
	\label{equ:zipf}
	P\{X = k\} = \frac{k^{-\alpha}}{\sum_{k=1}^{F} k^{-\alpha}}, ~ k = 1, 2, \ldots, F.
\end{equation}

\subsection{Cache Hit Rate}
\label{sec:Cache Hit Rate}

Let $\mathcal{M} = \left\{1, \ldots, m, \ldots, M\right\}$ be the set of $M$ oracles in DON, where $m$ is the $m$-th oracle.
Let $\mathcal{U} = \left\{1, \ldots, u, \ldots, U\right\}$ be the set of CRs, where $u$ is the $u$-th CR.
The TDC-Cache framework operates over $T$ time slots, denoted by $\mathcal{T} = \{1, \dots, t, \dots, T\}$, where $t$ is the $t$-th time slot.
\begin{definition}
	\label{def:Content Request Sequence}
	\textbf{(Content Request Sequence): } Let 
	$G_{{m}, {u}}^t = (g_{{m}, {u},1}^t, \ldots, g_{{m}, {u},k}^t,\ldots)$
	be an ordered and reproducible sequence of content requests from CR $u$ to oracle $m$ at time slot $t$, where $g_{{m}, {u},k}^t \in \mathcal{F}$ is the $k$-th request.
\end{definition}
From Definition \ref{def:Content Request Sequence}, oracle $m$ receives a sequence of content requests $G_{m}^t = \langle G_{m, 1}^t, G_{m, 2}^t, \ldots, G_{m, U}^t \rangle$ at time slot $t$, where $\langle \cdot \rangle$ is a concatenation operation.
Let $\boldsymbol{G}^t$ denote all the content requests $\langle G_{1}^t, G_{2}^t, \ldots, G_{M}^t \rangle$ received by all the oracles at time slot $t$.

Oracle $m$ maintains a cache pool with capacity $C_m$ to store frequently accessed content.
The cache pool state of oracle $m$ at time $t$ is denoted by $\mathcal{Y}_{m}^t=\left\{y_{m,1}^t, \dots, y_{m,j}^t, \dots, y_{m,C_{m}}^t\right\}$, where $y_{m,j}^t \in \mathcal{F}$ is the $j$-th content cached in $m$.
The total number of cached content is constrained by the cache capacity, i.e., $\lvert \mathcal{Y}_{m}^t \rvert \leq C_{m}$, where $\lvert \mathcal{Y}_{m}^t \rvert$ is the number of elements in $\mathcal{Y}_{m}^t$.

We define $h_{m,u,k}^t$ as a cache-hit indicator that indicates if the request $g_{{m}, {u}, k}^t$ belongs to $\mathcal{Y}_{m}^t$ (i.e., the cache-hit occurs):
\begin{equation}
	\label{equ:isHit}
	h_{m,u,k}^t = 
	\begin{cases} 
	1, & \text{if } g_{m, u, k}^t \in \mathcal{Y}_{m}^t, \\
	0, & \text{else.}
	\end{cases}
\end{equation}

To assess caching efficiency of TDC-Cache, we define the cache hit rate for oracle $m$ at time slot $t$ as
\begin{equation}
	\label{equ:hitrate}
	H_{m}^t = \frac{\sum_{u \in \mathcal{U}}\sum_{k=1}^{\lvert G_{{m}, {u}}^t \rvert} h_{m,u,k}^t}{\lvert G_{{m}}^t \rvert + \varepsilon},
\end{equation}
where $\varepsilon$ is a small constant to avoid division by zero, $\lvert G_{{m}}^t \rvert$ and $\lvert G_{{m}, {u}}^t \rvert$ are the number of requests in $G_{{m}}^t$ and $G_{{m}, {u}}^t$, respectively.
From (\ref{equ:hitrate}), the sum cache hit rate of all the oracles in the DON layer at time slot $t$ is
\begin{equation}
	\label{equ:global_hitrate}
	H_{\text{tot}}^t = \frac{\sum_{m\in\mathcal{M}}\sum_{u \in \mathcal{U}}\sum_{k=1}^{\lvert G_{{m}, {u}}^t \rvert} h_{m,u,k}^t}{\lvert \boldsymbol{G}^t \rvert + \varepsilon},
\end{equation}
where $\lvert \boldsymbol{G}^t \rvert$ is the total number of requests received by all the oracles in DON at time slot $t$.

\begin{lemma}
	\label{lemma:monotonicity}
	\textbf{(Monotonicity of Cache Hit Rate): }
	The cache hit rate $H_{\text{tot}}^t$ is a monotonically increasing function with respect to the cache hit rate $H_{m}^t, \forall m \in \mathcal{M}$.
\end{lemma}
\begin{IEEEproof}[Proof of Lemma \ref{lemma:monotonicity}]
	\label{proof:monotonicity}
	From (\ref{equ:global_hitrate}), we have
	\begin{equation}
	H_{\text{tot}}^t = \frac{\underset{m\in\mathcal{M}}{\sum}\left(\underset{u \in \mathcal{U}}{\sum} \overset{\lvert G_{{m}, {u}}^t \rvert}{\underset{k=1}{\sum}} h_{m,u,k}^t\right)}{\lvert \boldsymbol{G}^t \rvert + \varepsilon}.
    \end{equation}
	Let $ \text{Num}_{m} = \sum_{u \in \mathcal{U}}\sum_{k=1}^{\lvert G_{{m}, {u}}^t \rvert} h_{m,u,k}^t $, i.e., the total number of cache hits for $m$.
	Substituting this into (\ref{equ:global_hitrate}), we get
	\begin{equation}
		\label{equ:appendix_hitrate}
	H_{\text{tot}}^t = \frac{\sum_{m \in \mathcal{M}} \text{Num}_{m}}{|\boldsymbol{G}^t| + \varepsilon}.
    \end{equation}
	Using the definition of (\ref{equ:hitrate}), we have
	\begin{equation}
		\label{equ:Num_m}
	\text{Num}_{m} = H_{m}^t (|G_{m}^t| + \varepsilon).
    \end{equation}
	Substituting (\ref{equ:Num_m}) into (\ref{equ:appendix_hitrate}), we get
	\begin{equation}
	H_{\text{tot}}^t = \frac{\sum_{m \in \mathcal{M}} H_{m}^t (|G_{m}^t| + \varepsilon)}{|\boldsymbol{G}^t| + \varepsilon}.
    \end{equation}
	To compute $ \frac{\partial H_{\text{tot}}^t}{\partial H_{m}^t} $, we fix all $H_{m^\prime}^t$ for $m^\prime \neq m$ as constants.
	The derivative becomes
	\begin{equation}
	\frac{\partial H_{\text{tot}}^t}{\partial H_{m}^t} = \frac{\partial}{\partial H_{m}^t} \left( \frac{\left(\underset{m^\prime \in \mathcal{M}}{\sum} H_{m^\prime}^t + H_{m}^t \right) \left(|G_{m^\prime}^t| + \varepsilon\right)}{|\boldsymbol{G}^t| + \varepsilon} \right).
    \end{equation}
	Only the term with $H_{m}^t$ contributes to the derivative:
	\begin{equation}
	\frac{\partial H_{\text{tot}}^t}{\partial H_{m}^t} = \frac{|G_{m}^t| + \varepsilon}{|\boldsymbol{G}^t| + \varepsilon}.
    \end{equation}
	Since $|G_{m}^t| + \varepsilon > 0 $ and $ |\boldsymbol{G}^t| + \varepsilon > 0$, we have
	\begin{equation}
	\frac{\partial H_{\text{tot}}^t}{\partial H_{m}^t} > 0, ~ \forall m \in \mathcal{M}.
    \end{equation}
	Thus, $H_{\text{tot}}^t$ is a monotonically increasing function of $H_{m}^t$ for every oracle $ m $.
\end{IEEEproof}

At time slot $t$, oracle $m$ devises a caching strategy to maximize (\ref{equ:hitrate}) based on the content request sequence $G_{m}^t$ and the current cache pool $\mathcal{Y}_{m}^t$.
We denote this caching strategy by a matrix $A_{m}^t$ with size $\lvert G_{m}^t\rvert \times \lvert \mathcal{Y}_{m}^t \rvert$.
Each element $[A_{m}^t]_{k, j} \in \{0, 1\}$ in $A_{m}^t$ represents the decision to either evict or retain the $j$-th ($1 \leq j \leq \lvert \mathcal{Y}_{m}^t\rvert$) cached content in response to the $k$-th ($1 \leq k \leq \lvert G_{m}^t\rvert$) request of $G_{m}^t$.
Specifically, if $[A_{m}^t]_{k, j} = 1$, the outdated cached content is evicted; otherwise, the content is retained.
Moreover, the joint caching strategy at time slot $t$ is given by the union of individual oracle strategies: $\boldsymbol{A}^t = \underset{m \in \mathcal{M}}{\bigcup} A_{m}^{t}$, where $\underset{\cdot}{\bigcup}$ is the union operator.

\subsection{Content Retrieval Latency}
\label{sec:Content Retrieval Latency}

The content retrieval latency is defined duration between that the CR initiates a content request and the CR receives the content.
This latency varies based on whether the content is retrieved from the DON layer or the DLT layer.
Below are the two types of content retrieval latency models:

\subsubsection{Content retrieval latency of the DON layer}

If a cache hit occurs, the content is directly retrieved from the oracles.
The latency for the content request $g_{{m}, {u}, k}^t$ retrieved from DON at time slot $t$ is
\begin{equation}
	\label{equ:D_k}
	D_{g_{m, u, k}^t} = T_{g_{{m}, {u},k}^t}+\frac{C_{g_{{m}, {u},k}^t}}{\bar{v}},
\end{equation}
where $T_{g_{{m}, {u},k}^t}$ is the content addressing latency for $g_{{m}, {u}, k}^t$, $C_{g_{{m}, {u},k}^t}$ is the content size associated with $g_{{m}, {u},k}^t$, and $\bar{v}$ is the average transmission rate in the DLT layer.

\subsubsection{Content retrieval latency of the DLT layer}

If a cache miss occurs, the content is retrieved from the DLT layer.
This layer treats content as unstructured data (e.g., videos). It would be inefficient if this data were stored directly as a single and large file in a distributed storage network. Therefore, the content is partitioned into smaller and equal-sized binary large objects (BLOBs), i.e., the standard data type for raw binary bits, and is stored across different storage nodes \cite{zhao2022survey}.
To retrieve the whole content, all relevant BLOBs need to be collected and reassembled.
Therefore, the latency for content request $g_{{m}, {u}, k}^t$ retrieved from DLT at time slot $t$ is given by
\begin{equation}
	\label{equ:D_ds}
    D_{\text{ds}, g_{m, u, k}^t} = T_{g_{{m}, {u}, k}^t} + \left\lceil \frac{C_{g_{{m}, {u}, k}^t}}{C_{\text{chunk}}}\right\rceil \left(T_{\text{chunk}} + \frac{C_{\text{chunk}}}{\bar{v}}\right),
\end{equation}
where $T_{\text{chunk}}$ is the content addressing time for BLOB, $C_{\text{chunk}}$ is the size of each BLOB, and $\lceil \cdot \rceil$ is the ceiling function.

Overall, combining (\ref{equ:isHit}), (\ref{equ:D_k}), and (\ref{equ:D_ds}), the total content retrieval latency for all the requests at time slot $t$ is
\begin{equation}
	\label{equ:D_w}
	\hat{D}^t = \sum_{m \in \mathcal{M}} \sum_{u \in \mathcal{U}}\sum_{k=1}^{\lvert G_{{m}, {u}}^t\rvert} D_{g_{m, u, k}^t} h_{m,u,k}^t + D_{\text{ds}, g_{m, u, k}^t} \bar{h}_{m,u,k}^t, 
\end{equation}
where $\bar{h}_{m,u,k}^t = 1-h_{m,u,k}^t$ is the case of a cache miss.

\section{Latency Optimization of TDC-Cache}
\label{sec:Latency Optimization of TDC-Cache}

This section formulates the latency optimization problem of TDC-Cache and presents the proposed DRL-based solution.

\subsection{Problem Formulation}
\label{sec:Problem Formulation}
The content retrieval latency affects both the responsiveness of DApps and user experience.
This paper optimizes the caching strategy for each oracle to minimize the content retrieval latency.
In this context, we formulate the latency minimization problem as
\begin{align}
	\underset{\boldsymbol{A}^t}{\min} & \frac{1}{T}\sum _{t= 1}^{T} \hat{D}^t																			   \label{equ:problem}	\\
	 \text{s.t. } & \text{C1{:} }g_{{m}, {u}, k}^t\in \mathcal{F},~ \forall g_{{m}, {u}, k}^t \in G_{m, u}^t, \tag{\ref{equ:problem}{a}} \label{equ:problem_a}\\
	& \text{C2{:} }y_{m,j}^t\in \mathcal{F},~ \forall y_{m,j}^t \in \mathcal{Y}_{m}^t,								\tag{\ref{equ:problem}{b}} \label{equ:problem_b}\\
	& \text{C3{:} }y_{m,i}^t \neq y_{m,j}^t, ~ \forall y_{m,i}^t, y_{m,j}^t \in \mathcal{Y}_{m}^t, i \neq j,														\tag{\ref{equ:problem}{c}} \label{equ:problem_c}\\
	& \text{C4{:} }|\mathcal{Y}_{m}^t| \leq C_{m}, ~ \forall m \in \mathcal{M},					\tag{\ref{equ:problem}{d}} \label{equ:problem_d}
\end{align}
where $\boldsymbol{A}^t$ is the joint caching strategy of the oracle set $\mathcal{M}$ in DON at time slot $t$, C1 and C2 ensure that both requested content and cached items belong to the content set $\mathcal{F}$, C3 prevents the storage of duplicate content, and C4 constraints the total cache size at any time slot within the available storage capacity $C_{m}$.

To solve problem (\ref{equ:problem}), we need to optimize the caching decision for each time slot $t$.
However, as the number of contents or oracles increases, the complexity of problem solutions grows exponentially and has been proven to be NP-hard \cite{korbut2010exact}.
Given this intractability, we explore a DRL-based approach inspired by \cite{ma2024efficient}, which enables distributed cooperative agents to dynamically learn near-optimal policies.
These policies are derived from the complete sequence of actions taken by the agents and their corresponding observations from the initial to the current time slot.

\subsection{Decentralized Caching Strategy}
\label{sec:The Implementation of DRL-DC}
As described in Section \ref{sec:The Architecture of TDC-Cache}, the DON layer comprises a group of oracles that independently make caching decisions without centralized coordination or access to the state of other oracles.
Given these characteristics, this paper formulates problem (\ref{equ:problem}) as a Decentralized Partially Observable Markov Decision Process (Dec-POMDP)\cite{oliehoek2016concise}.
Specifically, the Dec-POMDP model is defined by the tuple $\langle \mathcal{M}, \mathcal{S}, \mathcal{O}, \mathcal{A}, P, Z, r, \gamma \rangle$, 
where $\mathcal{M}$ is a set of $M$ oracles, $\mathcal{S}$, $\mathcal{O}$, and $\mathcal{A}$ are the state, observation, and action spaces, respectively.
Moreover, $P(\boldsymbol{s}^{t+1} | \boldsymbol{s}^t, \boldsymbol{a}^t)$ is the probability of transitioning from state $\boldsymbol{s}^t$ to state $\boldsymbol{s}^{t+1}$ given the action $\boldsymbol{a}^t$,
$Z(\boldsymbol{o}^{t+1} | \boldsymbol{a}^t, \boldsymbol{s}^{t+1})$ is the probability of observing observations $\boldsymbol{o}^{t+1}$,
$r(\boldsymbol{s}^t, \boldsymbol{a}^t)$ is the immediate reward obtained for taking action $\boldsymbol{a}^t$ in state $\boldsymbol{s}^t$,
and $0 \leq \gamma \leq 1$ is the discount factor for future rewards.

\subsubsection{Observation Space}
At time slot $t$, the observation at oracle $m$, defined as $O_{m}^t = \{\mathcal{Y}_{m}^t, G_{m}^t, C_{m}\}$, includes the current state of the cache pool, the content request sequence, and the cache pool capacity.
Further, the state of DON at time slot $t$ is $\boldsymbol{S}^t=\{O_{1}^t, \ldots, O_{M}^t\}$, which collects the observations from all the oracles in $\mathcal{M}$.

\subsubsection{Action Space}
To maximize the cache hit rate, oracle $m$ optimizes the caching strategy $A_{m}^t$ by dynamically analyzing the content request sequence $G_{m}^t$ and the cache pool $\mathcal{Y}_{m}^t$ under a cache capacity limit of $C_{m}$.
Similarly, in DRL, an agent selects actions based on observation $O_{m}^t$ to maximize its reward.
Therefore, we consider the caching strategy of each oracle as the action taken by an individual agent, while the joint caching strategy of all oracles corresponds to the joint action in DRL.

\subsubsection{Reward Function}
The reward function at time slot $t$ is
\begin{equation}
	\label{equ:global_reward}
	r^t=\sum_{m \in \mathcal{M}} r_{m}^t,
\end{equation}
where $r_{m}^t$ is the reward function for oracle $m$ at time slot $t$, defined as
\begin{equation}
	\label{equ:local_reward}
	r_{m}^t=\omega I_{m}^t+\lambda \xi_{m}^t,
\end{equation}
where $I_{m}^t$ and $\xi_{m}^t$ are the incentive and penalty components respectively, $\omega \geq 1$ is a weighting coefficient, and $0 \leq \lambda \leq 1$ is a balancing factor.
Specifically, the incentive component of (\ref{equ:local_reward}) is
\begin{equation}
I_{m}^t=h_{m}^t-h_{m}^{t-1},
\end{equation}
where $h_{m}^t = \sum_{u \in \mathcal{U}}\sum_{k=1}^{\lvert G_{{m}, {u}}^t \rvert} h_{m,u,k}^t$ represents the number of caching hit for oracle $m$ at time slot $t$.
For completeness, we initialize $h_{m}^0 = 0$.
This component highlights the fluctuations in cache hit counts, thereby enabling each oracle to rapidly adjust its caching strategy in response to the dynamic content popularity.
The penalty component of (\ref{equ:local_reward}) is given by
\begin{equation}
	\xi_{m}^t = \mu \xi_{m, 1}^t + \rho \xi_{m, 2}^t,
\end{equation}
where $\xi_{m, 1}^t$ and $\xi_{m, 2}^t$ are the penalties for violating constraints (\ref{equ:problem_c}) and (\ref{equ:problem_d}) by oracle $m$ at time slot $t$ respectively, and $\mu \leq 0$ and $\rho \leq 0$ are penalty hyperparameters.
Moreover, $\xi_{m, 1}^t = \sum_{i=1}^{|\mathcal{Y}_{m}^t|} \sum_{\substack {j=1 \\ i \neq j}}^{|\mathcal{Y}_{m}^t|} \mathbf{1}_{ \{ y_{m,i}^t = y_{m,j}^t \} }$, where $\mathbf{1}_{\{\cdot\}}$ is an indicator function that indicates if $y_{m,i}^t = y_{m,j}^t$, and $\xi_{m, 2}^t = (|\mathcal{Y}_{m}^t| - C_{m})^2$ implies that both underutilization and overutilization of cache capacity result in penalties. Notably, the underutilization is potentially caused by inefficient caching strategies and inevitably leads to low cache hit rates.

\begin{definition}
	\label{def:Action-Observation History}
	\textbf{(Action-Observation History): } 
	The action-observation history of oracle $m$ at time slot $t$ is defined as
	\begin{equation}
		\tau_{m}^t = \left\{ o_{m}^1, a_{m}^1; o_{m}^2, a_{m}^2; \dots; o_{m}^{t-1}, a_{m}^{t-1}\right\},
	\end{equation}
	where $o_{m}^{t-1}$ and $a_{m}^{t-1}$ are the observed values for $O_{m}^{t-1}$ and $A_{m}^{t-1}$, respectively.
\end{definition}
With Definition \ref{def:Action-Observation History}, let $\boldsymbol{\tau}^t = \left\{\tau_{1}^t, \ldots, \tau_{M}^t\right\}$ be the joint action-observation history of all the oracles at time slot $t$.

\begin{theorem}
	\label{theorem:consistency}
	\textbf{(Consistency between Decentralized and Centralized Policies): }In a distributed cooperative environment, consistency between decentralized and centralized policies holds if the locally optimal actions, derived independently by each oracle based on its respective action-observation history, are identical to the globally optimal joint actions derived from the joint action-observation history of all the oracles:
	\begin{equation}
		\underset{\boldsymbol{a}^t}{\arg \max}~H_{\textup{tot}}^t(\boldsymbol{\tau}^t, \boldsymbol{a}^t)=\bigcup_{m \in \mathcal{M}} \underset{a^t_m}{\arg \max}~H_m^t(\tau^t_m, a^t_m),
	\end{equation}
	where $\boldsymbol{a}^t$ is the joint action of all the oracles at time slot $t$, $H_{\textup{tot}}^t(\boldsymbol{\tau}^t, \boldsymbol{a}^t)$ has been defined as the global cache hit rate at time slot $t$ in (\ref{equ:global_hitrate}), and $H_m^t(\tau^t_m, a^t_m)$ is the cache hit rate for oracle $m$ at time slot $t$ based on the action-observation history $\tau^t_m$ and action $a^t_m$.
\end{theorem}
\begin{IEEEproof}[Proof of Theorem \ref{theorem:consistency}]
	Let ${a}_{m}^{t, *} = \underset{a_{m}^{t}}{\arg \max}~H_m^t (\tau_m^t,a_m^t)$ and ${\boldsymbol{a}}^{t, *} = \underset{\boldsymbol a^t}{\arg\max}~H_{\mathrm{tot}}^t(\boldsymbol{\tau}^t,\boldsymbol a^t)$ denote the optimal local action for the oracle $m$ and the optimal joint actions for the oracles of DON at time slot $t$, respectively.
	According to Lemma \ref{lemma:monotonicity}, $H_{\mathrm{tot}}^t(\boldsymbol{\tau}^t,\boldsymbol a^t)$ is a monotonically increasing function of each $H_m^t(\tau_m^t,a_m^t)$, satisfying $\frac {\partial H_\mathrm{tot}} {\partial H_{m}} > 0,~ \forall m \in \mathcal{M}$.
	% H_{\mathrm{tot}}^t(\boldsymbol{\tau}^t,\boldsymbol a^{t, *}) & = 
	Thus, for any time slot $t$, the following inequalities hold
	\begin{align}
		\label{equ:appendix_inequality}
		& f\left(H_{1}^t(\tau_{1}^t,A_{1}^t), \ldots, H_{M}^t(\tau_{M}^t,A_{M}^t)\right) \nonumber \\
		& \leq f\left(H_{1}^t(\tau_{1}^t,a_{1}^{t, *}), \ldots, H_{M}^t(\tau_{M}^t,A_{M}^t)\right) \nonumber \\
		& \ldots \nonumber \\
		& \leq f\left(H_{1}^t(\tau_{1}^t,a_{1}^{t, *}), \ldots, H_{M}^t(\tau_{M}^t,a_{M}^{t, *})\right),
	\end{align}
	where $f(\cdot)$ is a monotonically increasing function in each argument.
	From (\ref{equ:appendix_inequality}), we define
	\begin{align}
		\label{equ:appendix_equality}
		H_{\mathrm{tot}}^t(\boldsymbol{\tau}^t,\boldsymbol a^{t, *}) = f\left(H_{1}^{t}(\tau_{1}^t,a_{1}^{t, *}), \ldots, H_{M}^{t}(\tau_{M}^t,a_{M}^{t, *})\right),
	\end{align}
	where it implies that any joint action formed by the locally optimal actions of all the oracles is globally optimal. Conversely, if $\boldsymbol a^{t,*}$ is globally optimal but one of its components is not locally optimal, then substituting that component with the locally optimal action would strictly increase $f$, which contradicts the global optimality.
	Therefore, the optimal global action yields 
	\begin{equation}
	\arg\max_{\boldsymbol a^t} H_{\mathrm{tot}}^t(\boldsymbol{\tau}^t,\boldsymbol a^t)
	= \bigcup_{m\in\mathcal M} \arg\max_{a_m^t} H_m^t(\tau_m^t,a_m^t),
    \end{equation}
	which proves Theorem \ref{theorem:consistency}.
\end{IEEEproof}

This paper uses QMIX, a value decomposition-based multi-agent DRL algorithm, to solve the Dec-POMDP problem \cite{rashid2020monotonic}.
Notably, the usage of QMIX is required to satisfy monotonicity of value decomposition and consistency of action selection.
In what follows, we show that the proposed caching strategy satisfies the usage requirement of QMIX.
First, Lemma \ref{lemma:monotonicity} verifies that the proposed decentralized caching strategy guarantees monotonicity in individual contributions to the global cache hit rate.
Second, Theorem \ref{theorem:consistency} validates the consistency between decentralized and centralized policies.

The QMIX training process builds on the standard Deep Q-Learning (DQN) algorithm and minimizes a loss function:
\begin{equation}
	\label{equ:loss}
	\mathcal{L}(\theta)= \sum_{i = 1}^{b} \left[ \left( y_i^{\text{tot}} - Q_{\text{tot}}(\boldsymbol{\tau}^t, \boldsymbol{s}^t, \boldsymbol{a}^t; \theta) \right)^2 \right],
\end{equation}
where $b$ is the batch size of transitions sampled from the replay buffer, and $Q_{\text{tot}}(\boldsymbol{\tau}^t, \boldsymbol{s}^t, \boldsymbol{a}^t; \theta)$ is the total $Q$-value parameterized by $\theta$ for the independent $Q$-network.
Moreover, the DQN target is given by
\begin{equation}
y_{i}^{\text{tot}} = r^t + \gamma \max_{\boldsymbol{a}^{t+1}} Q_{\text{tot}}(\boldsymbol{\tau}^{t+1}, \boldsymbol{s}^{t+1}, \boldsymbol{a}^{t+1}; \theta^\prime), 
\end{equation}
where $r^t$ is the global reward calculated by (\ref{equ:global_reward}), $\gamma$ is the discount factor that balances immediate and future rewards, $\boldsymbol{\tau}^{t+1}$, $\boldsymbol{a}^{t+1}$, and $\boldsymbol{s}^{t+1}$ are the next action-observation history, joint action, and state respectively, and $\theta^\prime$ is the parameters of the target network.

\section{Proof of Cooperative Learning}
\label{sec:Proof of Cooperative Learning}

\begin{figure*}[!t]
	\centering
	\includegraphics[width=1.0\textwidth]{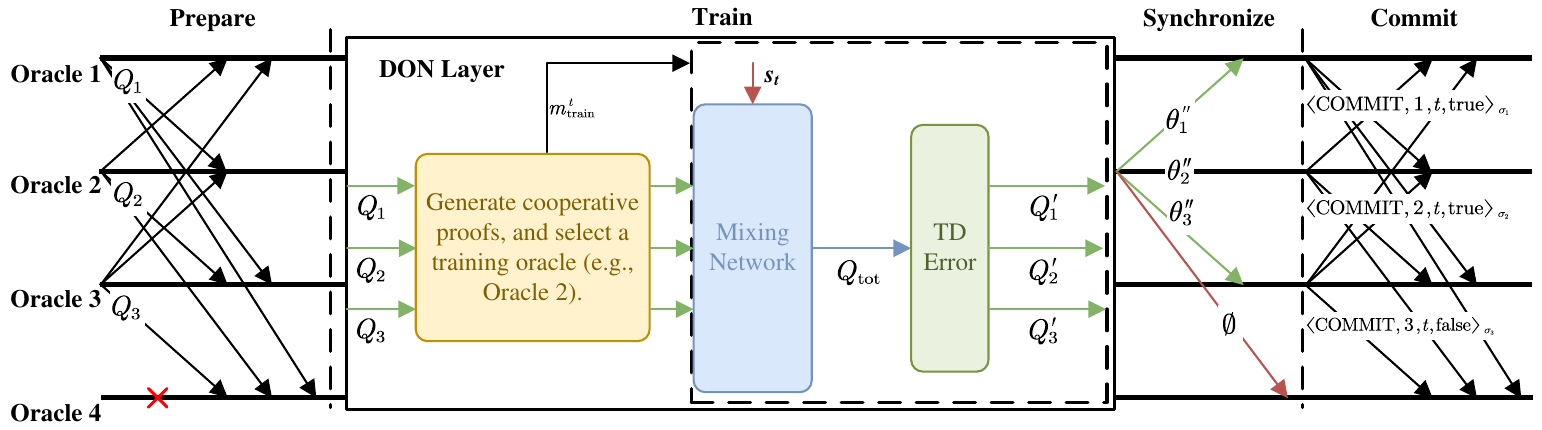}
	\caption{Four phases of PoCL.}
	\label{fig:PoCL}
\end{figure*}

In a fully decentralized Dec-POMDP setting, malicious oracles have motives to manipulate data for personal gain \cite{guerraoui2024byzantine}.
To address this, we propose Proof of Cooperative Learning (PoCL), a Byzantine fault tolerance mechanism to protect against malicious oracles from destructing the consistency of DRL-DC strategy.
Inspired by Proof of Useful Work (PoUW) \cite{ball2017proofs} and Practical Byzantine Fault Tolerance (PBFT) \cite{castro1999practical}, PoCL promotes both the trustworthiness of $Q$-value aggregation and the accuracy of honest training oracle selection in TDC-Cache.
As shown in Fig. \ref{fig:PoCL}, PoCL achieves global consensus over four sequential phases: \textbf{Prepare}, \textbf{Train}, \textbf{Synchronize}, and \textbf{Commit}.
These phases are designed to ensure secure and efficient consensus in a fully decentralized and adversarial network. Specifically, the first two phases are used to select a training oracle by a global consensus within the active oracles, while the latter two phases disseminate and validate the caching strategy update proposed by the training oracle.

\subsection{Prepare Phase}
During \textbf{Prepare}, each oracle $m \in \mathcal{M}$ broadcasts a \textit{prepare} message, denoted by $\langle \text{PREPARE}, m, t, Q_{m}, r_{m}, \rangle_{\sigma_{m}}$, where $t$ is the current time slot, $Q_{m}$ is the $Q$-value of QMIX, and $r_{m}$ has been defined in (\ref{equ:local_reward}).
This message is signed by a valid signature verifiable $\sigma_{m}$.

The oracle then waits for a period of $T_\text{pre}$ to receive \textit{prepare} messages from others.
Upon receipt, it verifies each message's integrity and if it belongs to the current time slot $t$.
After verification, valid messages are added to the oracle's message pool $\mathcal{M}_\text{pre}^t \subset \mathcal{M}$, while invalid ones are discarded.
At the end of $T_\text{pre}$, the oracle stops accepting messages.

\subsection{Train Phase}
During this phase, the oracles in DON first evaluate the cooperative proofs of all the participating oracles and then select a training oracle to train the decentralized caching strategy.
Specifically, we design the cooperative proof of oracle $m$ at time slot $t$ as $\phi_{m}^t = \{\psi_m^t, r_{m}^t\}$, where $\psi_m^t$ is the hit count computed from the value function $Q_m$ of oracle $m$ on a shared test set.
To ensure consistency, this test set is pseudo-randomly reorganized based on content request sequence $\boldsymbol{G}^{t-1}$ and then delivered to each oracle via OMC.

Using the cooperative proofs, let us specify the selection criterion of the training oracle.
Suppose that the training oracle at time slot $t$ is $m_{\text{train}}^t \in \mathcal{M}_\text{pre}^t$.
First, to evaluate the cooperative proof of oracle $m$, we define the average distance between its cooperative proof and those of other oracles in the message pool $\mathcal{M}_\text{pre}^t$ as the evaluation metric:
\begin{equation}
	\label{equ:likelihood}
	L_m^t = \frac{\beta^{-\varpi_m^t }}{\lvert\mathcal{M}_\text{pre}^t\rvert-1}\sum_{\overset{m^\prime \in \mathcal{M}_\text{pre}^t}{m^\prime \neq m}}l(\phi_m^t, \phi_{m^\prime}^t),
\end{equation}
where $0 < \beta \leq 1$ is a decay factor, $\lvert \mathcal{M}_\text{pre}^t \rvert$ is the number of \textit{prepare} messages received by oracle $m$ at time slot $t$, $l(\phi_m^t, \phi_{m^\prime}^t) = \sqrt{(\psi_m^t - \psi_{m^\prime}^t)^2 + (r_m^t - r_{m^\prime}^t)^2}$ is the Euclidean distance of the cooperative proofs between oracle $m$ and $m^\prime$.
Second, to prevent the same oracle from being repeatedly selected, we introduce a history-based adjustment factor $\varpi_m^t$ to record the number of consecutive time slots oracle $m$ has served as a training oracle:
\begin{equation}
	\label{equ:varpi}
	\varpi_m^t = \begin{cases}
		\varpi_m^{t-1} + 1, & \text{if } m = m_{\text{train}}^{t - 1}, \\
		0, & \text{else}.
	\end{cases}
\end{equation}
Finally, let $\mathbb{L}^t = \{L_1^t, \ldots, L_{\lvert \mathcal{M}_\text{pre}^t \rvert}^t\}$ be the set of average distances for all oracles in the message pool $\mathcal{M}_\text{pre}^t$ at time slot.
In this context, the training oracle is identified as the oracle with the smallest average distance:
\begin{equation}
	\label{equ:m_train}
	m_{\text{train}}^t= \underset{m \in \mathcal{M}_\text{pre}^t}{\arg \min}~\mathbb{L}^t.
\end{equation}
Each oracle in DON executes the selection process independently.
However, since $\mathbb{L}^t$ is computed deterministically, all honest oracles reach the same training oracle selection result, which guarantees the consistency of training oracle selection in DON.

In this phase, the training oracle is required to allocate computational resources to train the caching strategy model, and generates a global value function by mixing $Q$-values from all the oracles in $\mathcal{M}_\text{pre}^t$:
\begin{equation}
	\label{equ:mixing_network}
	Q_{\text{tot}}(\tau,a) = f\left(Q_{1}(\tau_{1},a_{1}),\ldots,Q_{M}(\tau_{M}, a_{M});\boldsymbol{\theta}\right),
\end{equation}
where $f$ is the mixing function of QMIX and $\boldsymbol{\theta}$ is the mixing network parameters.
Subsequently, $m_{\text{train}}^t$ updates the mixing network and local $Q$-network based on (\ref{equ:loss}).

For the sake of time efficiency, we stipulate that the training process of training oracle must be completed within $T_\text{tra}$ time.
Otherwise, the oracle will be marked as a faulty oracle by other oracles.

\subsection{Synchronize Phase}

The training oracle $m_{\text{train}}^t$ sends updated parameters $\theta_{m}^{\prime}$ to corresponding oracles via a signed \textit{sync} message, defined as $\langle \text{SYNC}, m_{\text{train}}^t, t, \theta_{m}^{\prime \prime} \rangle_{\sigma_{m_{\text{train}}^t}}$, where $\theta_{m}^{\prime \prime}$ is $\theta_{m}^{\prime}$ encrypted with the public key of oracle $m$.
The \textit{sync} message is signed using $m_{\text{train}}^t$'s private key for integrity verification.
Meanwhile, $m_{\text{train}}^t$ sends a \textit{null} message $\langle \text{SYNC}, m_{\text{train}}^t, t, \emptyset \rangle_{\sigma_{m_{\text{train}}^t}}$ to unavailable oracles (i.e., present in $\mathcal{M}$ but absent in $\mathcal{M}_\text{pre}^t$).
These unavailable oracles may include honest oracles that simply missed the \textbf{Prepare} phase.
Therefore, the \textit{null} message is used to inform these unavailable oracles that the \textbf{Prepare} and \textbf{Train} phases have completed, prevent them from waiting indefinitely, and ultimately maintain consensus liveness.

\subsection{Commit Phase}

Upon receiving \textit{sync} message, oracle $m$ verifies the signature using $m_{\text{train}}^t$'s public key and decrypts $\theta_{m}^{\prime \prime}$ with its private key $SK_{m}$ to obtain $\theta_{m}^{\prime}$.
It then evaluates $\theta_{m}^{\prime}$ using the shared test set, downloaded during the \textbf{Train} phase, to compute $\psi_m^{\prime t}$.
If $\psi_m^{\prime t} > \psi_m^t$, the result is considered improved, and the oracle broadcasts $\langle \text{COMMIT}, m, t, \text{true} \rangle_{\sigma_m}$; otherwise, it broadcasts $\langle \text{COMMIT}, m, t, \text{false} \rangle_{\sigma_m}$.

Upon receiving a \textit{commit} message, the oracle authenticates its signature and confirms that the sender belongs to $\mathcal{M}_\text{pre}^t$.
If more than $2 N_{\text{b}}$ valid \textit{commit} messages (excluding itself) are received, and at least $N_{\text{b}} + 1$ of them (including itself) are $\langle \text{COMMIT}, m, t, \text{true} \rangle_{\sigma_m}$, the local model is updated to $\theta_{m}^{t+1} = \theta_{m}^{\prime}$, where $N_{\text{b}}$ is the number of Byzantine Oracle.
Otherwise, the training oracle is suspected to be Byzantine, and the update is discarded.
The \textbf{Commit} phase then concludes, and TDC-Cache advances to time slot $t+1$.

\begin{remark}
	Unlike PBFT where the primary node remains invariant unless it fails or is voted out, PoCL dynamically selects a training oracle in each round based on the cooperative proof.
	This approach eliminates the need for view changes.
	Consequently, PoCL not only enhances the security and decentralization of DON by leveraging cooperative proofs to ensure the selection of reliable training oracle, but also improves the efficiency of DRL-DC by mitigating computational redundancy and reducing resource consumption.
\end{remark}

\section{Performance Analysis of Consensus}
\label{sec:Performance Analysis of Consensus}

This section analyzes the security, liveness, and success consensus rate of PoCL.
Suppose there are $N_\text{b}$ Byzantine oracles, $N_\text{f}$ faulty oracles, and $N_\text{h} = N_\text{b} + 1$ honest oracles.
Byzantine oracles can disrupt decentralized caching decisions by sharing false or inconsistent information, while faulty oracles cannot function normally due to software or hardware problems \cite{okegbile2022practical}.

\subsection{Safety}
\label{sec:Safety}

We study PoCL's security in terms of Byzantine fault tolerance and decentralization.

\subsubsection{Byzantine Fault Tolerance}

The PoCL consensus maintains the consistency of decentralized caching decisions even if the selected training oracle is Byzantine.
Due to the spatio-temporal locality of cached content \cite{breslau1999web}, cooperative proofs from honest oracles tend to be closely clustered, and their Euclidean distances remain small.
Consequently, the training oracle $m_{\text{train}}^t$ is more likely to be selected from the honest oracles when $N_\text{h} > N_\text{b}$.
Even if a Byzantine oracle is chosen as the training oracle and attempts to manipulate the caching strategy (e.g., reducing the cache hit rate), PoCL prevents malicious updates from being adopted.
This is because the training oracle’s broadcasted caching model is required to receive at least $N_\text{b} + 1$ valid $\langle \text{COMMIT}, m, t, \text{true} \rangle_{\sigma_m}$ messages.
However, honest oracles detect the received caching model under the model update rules defined in \textbf{Commit} and reject the malicious caching model by broadcasting $\langle \text{COMMIT}, m, t, \text{false} \rangle_{\sigma_m}$.
Consequently, Byzantine training oracle fails to obtain the required $N_\text{b} + 1$ valid $\langle \text{COMMIT}, m, t, \text{true} \rangle_{\sigma_m}$ messages.
\subsubsection{Mitigation of Centralization Risk}
PoCL mitigates centralization risks by reducing the likelihood of repeatedly selecting the same oracle as the training oracle \cite{altamimi2023not}.
Such centralization may lead to a single point of failure and undermine the fairness of distributed consensus \cite{shi2024blockchain, zhu2025randomized}.
To address this, PoCL uses a history-based decay factor $\varpi_m^t$ in (\ref{equ:varpi}) to increase the average Euclidean distance of frequently selected oracles and to lower their future selection probability.
Consequently, PoCL enhances the decentralization of DON.
\subsubsection{Mitigation of Collusion Attacks}
PoCL also mitigates collusion attacks targeting the training oracle selection process.
The historical adjustment factor, computed from immutable on-chain records, prevents colluding oracles from manipulating their selection probabilities within a single slot.
Even if colluders succeed in electing a malicious training oracle by submitting falsified cooperative proofs \cite{xiao2022sca}, the elected oracle is required to complete the training task; otherwise, its model update would fail to obtain sufficient votes during the \textbf{Commit} phase.
In subsequent slots, $\varpi_m^t$ further decreases the chance of the same malicious oracle being selected again.
Therefore, as long as the number of colluding oracles is below $N_b$, PoCL ensures the fairness and security of the training oracle selection.

\subsection{Liveness}
\label{sec:Liveness}
Liveness ensures that TDC-Cache maintains the consistency of decentralized caching decisions among oracles within a limited period.
Even under network latency, oracle failures, or malicious attacks, TDC-Cache continues to operate normally and eventually restores consistency.
In the previous subsection, security is ensured as long as at least $N_\text{b}+1$ honest oracles exist among $M$, which also guarantees liveness.
Suppose that TDC-Cache is initially composed of $N_\text{h} + N_\text{f}$ honest oracles and $N_\text{b}$ Byzantine oracles.
Over time, both honest and Byzantine oracles may become faulty.
This part analyzes the worst-case scenario where the training oracle $m_{\text{train}}^t$ becomes unavailable.
If $m_{\text{train}}^t$ fails to send the \textit{sync} message during the $T_\text{tra}$, other oracles enter a waiting state.
If a \textit{sync} message is received within $T_\text{tra}$, the consensus proceeds normally.
Otherwise, the oracle with the next smallest average distance in $\mathcal{M}_\text{pre}^t$ is selected as $m_{\text{train}}^t$.

This selection process repeats at most $N_\text{f}$ times.
If no \textit{sync} message is received after $N_\text{f}$ attempts, it is assumed that the number of faulty oracles exceeds $N_\text{b} + 1$, and the consensus process is considered failed.
Thus, the consensus process ensures liveness as long as $N_\text{f} < N_\text{b} + 1$.

\subsection{Success Consensus Rate}
The success consensus rate is defined as the probability that all phases of PoCL complete successfully without failure.
This metric evaluates both the liveness and reliability of PoCL in the presence of faulty or Byzantine oracles. Accordingly, the decentralized network with a higher consensus success rate exhibits greater stability.
In practice, each phase of PoCL has a probability of failure $P_\text{f}$, with failures occurring independently across phases.
In this context, we investigate the success probability of each phase to characterize the overall reliability of PoCL.

\subsubsection{Success Rate of the Prepare Phase}

During the \textbf{Prepare} phase, each oracle independently broadcasts \textit{prepare} messages to others.
Consensus fails if the number of faulty oracles exceeds $N_\text{f}$, as demonstrated in Section \ref{sec:Safety}.
Therefore, the success probability in this phase is
\begin{equation}
	\label{equ:prepare_success_rate}
	P_\text{pre} = \sum_{i=0}^{N_\text{f}} {M-N_\text{b} \choose i} P_\text{f}^i(1-P_\text{f})^{M-N_\text{b}-i},
\end{equation}
where ${M-N_\text{b} \choose i}$ is the number of ways to select $i$ faulty oracles from $M-N_\text{b}$ non-Byzantine oracles.

\subsubsection{Success Rate of the Train and Synchronize Phases}

Oracles operate independently without communications in the \textbf{Train} phase.
Their states (i.e., active or failed) only become apparent during the \textbf{Synchronize} phase.
Consequently, the \textbf{Train} and \textbf{Synchronize} phases share the same success probability.
If $i$ oracles fail in \textbf{Prepare}, $M - i$ active oracles proceed to \textbf{Train} to select $m_{\text{train}}^t$ and broadcast \textit{sync} messages.
If the training oracle fails, a new ${m^\prime}_{\text{train}}^t$ is selected.
Therefore, at most $N_\text{f}-i$ additional oracles fail during these phases.
The success probability in this phase is
\begin{equation}
	\label{equ:train_success_rate}
	P_\text{tra} = \sum_{j=0}^{N_\text{f}-i}P_\text{f}^j(1-P_\text{f}).
\end{equation}

\subsubsection{Success Rate of the Commit Phase}

If $i$ oracles fail during \textbf{Prepare} and $j$ oracles fail during \textbf{Train} and \textbf{Synchronize}, $M - i - j$ oracles remain active in \textbf{Commit} to broadcast \textit{commit} messages.
At this phase, at most $N_\text{f}-i-j$ additional oracles fail.
The success probability in this phase is
\begin{equation}
	\label{equ:commit_success_rate}
	P_\text{com} = \sum_{k=0}^{N_\text{f}-i-j} {M-N_\text{b}-i-j \choose k} P_\text{f}^k(1-P_\text{f})^{M-N_\text{b}-i-j-k}.
\end{equation}

Based on the analysis above, we obtain the consensus success probability of PoCL as
\begin{equation}
	\label{equ:consensus_success_rate}
	P_\text{suc} = P_\text{pre}P_\text{tra}P_\text{com}.
\end{equation}
This probability (\ref{equ:consensus_success_rate}) quantitatively evaluates liveness and reliability of PoCL by adjusting the total number of oracles $M$, the number of Byzantine oracles $N_\text{b}$, the faulty oracle tolerance threshold $N_\text{f}$, and the individual failure probability $P_\text{f}$.

\section{Experimental Results}
\label{sec:Experimental Results}

In this section, we present the experimental results of TDC-Cache, including the success rate of PoCL, the cache hit rate of DRL-DC, and the content retrieval latency.
The experiments are conducted on a system running Ubuntu 24.04 LTS, equipped with an Intel Core i9-14900K CPU, NVIDIA GeForce RTX 4090 GPU with 24 GB of VRAM, 128 GB of RAM, and a 1TB SSD.

\subsection{Success Consensus Rate}
We conduct the experiment of PoCL in Python 3.10.
The experiment uses two parameters: the number of oracles $M$ and the probability of failure $P_\text{f}$.
With a random seed $2^{64}-1$, the experiment models four phases where oracles may move from an honest or Byzantine state to a faulty state.
This experiment initializes with $N_\text{b} = \lfloor \frac{M-1}{3} \rfloor$ Byzantine oracles and $M-N_\text{b}$ honest oracles, where $\lfloor \cdot \rfloor$ is the floor function.
At the end of PoCL, the experiment counts responses $\langle \text{COMMIT}, m, t, \text{true} \rangle_{\sigma_m}$ and $\langle \text{COMMIT}, m, t, \text{false} \rangle_{\sigma_m}$ separately, and determines consensus success or failure based on the following conditions:
\begin{enumerate}
	\item Consensus fails if the number of active oracles reduces to zero at any phase.
	\item If $m_{\text{train}}^t$ is a Byzantine oracle, consensus succeeds if the number of responses $\langle \text{COMMIT}, m, t, \text{false} \rangle_{\sigma_m}$ exceeds $N_\text{b}$; otherwise, it fails.
	\item If $m_{\text{train}}^t$ is an honest oracle, consensus succeeds if the number of responses $\langle \text{COMMIT}, m, t, \text{true} \rangle_{\sigma_m}$ exceeds $N_\text{b}$; otherwise, it fails.
	\item All other cases result in consensus failure.
\end{enumerate}
For statistical analysis, we conduct the experiment 10,000 times under the same $M$ and $P_\text{f}$ to compute the experimental result of success consensus rate $P_\text{suc}$. 
Additionally, we use these conditions to compute the theoretical result of $P_\text{suc}$ based on (\ref{equ:consensus_success_rate}).

\begin{figure}[!t]
	\centering
	\includegraphics[width=2.5in]{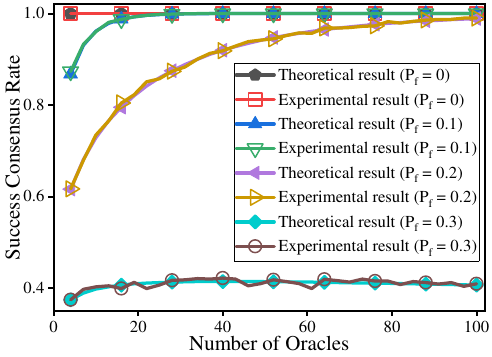}
	\caption{Comparison of experimental and theoretical results under different $M$ and $P_\text{f}$.}
	\label{fig:Comparison_Simulation_Analysis}
\end{figure}

Fig. \ref{fig:Comparison_Simulation_Analysis} compares theoretical and experimental results of success consensus rate $P_\text{suc}$ for different failure probabilities $P_\text{f}$.
Each curve corresponds to a failure probability $P_\text{f} \in \{0, 0.1, 0.2, 0.3\}$.
It is observed that when $P_\text{f} = 0$, both the theoretical and experimental results yield $P_\text{suc} = 1.0$.
When $P_\text{f} = 0.1$ and $P_\text{f} = 0.2$, $P_\text{suc}$ converges to 1.0 as $M$ increases.
However, when $P_\text{f} = 0.3$, $P_\text{suc}$ remains relatively low.
These observations further corroborate the consistency between the experimental and theoretical results.

\begin{figure}[!t]
	\centering
	\includegraphics[width=2.5in]{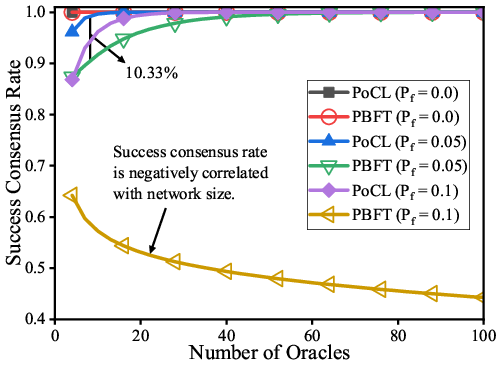}
	\caption{Comparison of success consensus rates between PoCL and PBFT under different $M$ and $P_\text{f}$ conditions.}
	\label{fig:Comparison_PBFT_PoCL}
\end{figure}

Fig. \ref{fig:Comparison_PBFT_PoCL} compares success consensus rate $P_\text{suc}$ between PoCL and PBFT of \cite{luo2023performance} under failure probabilities $P_\text{f} \in \{0, 0.05, 0.1\}$.
It is observed that when $P_\text{f} = 0$, both PoCL and PBFT achieve $P_\text{suc} = 1.0$.
When $P_\text{f} = 0.05$, both $P_\text{suc}$ gradually go up as the number of oracles and approaches 1.0, with PoCL consistently outperforming PBFT by up to 10\%.
However, the success rate of PBFT declines as the number of oracles increases at $P_\text{f} = 0.1$, whereas PoCL continues to improve and eventually reaches 1.0.

\begin{figure}[!t]
	\centering
	\includegraphics[width=2.5in]{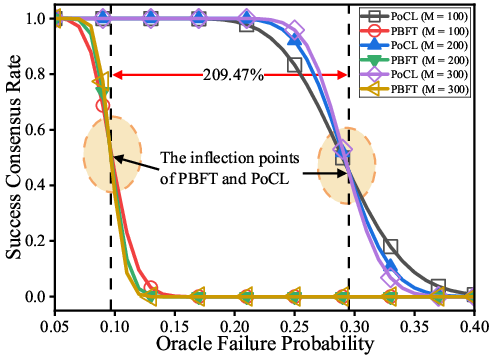}
	\caption{Impact of oracle failure probability on success consensus rate for PoCL and PBFT.}
	\label{fig:Change_Threshold}
\end{figure}

Fig. \ref{fig:Change_Threshold} shows the impact of failure probability $P_\text{f}$ on the success consensus rates $P_\text{suc}$ of PoCL and PBFT across network sizes $M \in \{100, 200, 300\}$.
First, it is observed that $P_\text{suc}$ are negatively correlated with the failure probability $P_\text{f}$ in both PBFT and PoCL.
The success rate of PBFT drops sharply when $P_\text{f} > 0.07$, whereas PoCL maintains a higher success rate when $P_\text{f} < 0.20$ before starting to decline.
Second, within $P_\text{f} \in [0.15, 0.25]$, the success rate of PBFT approaches 0, while PoCL maintains a high success rate close to 1.0.
Last, both algorithms exhibit inflection points, beyond which an increase in network size leads to a decrease in success rates.
However, PoCL’s inflection point is 209.47\% higher than that of PBFT.
Notably, the performance gap between PoCL and PBFT widens as the network size expands.
These observations demonstrate superior fault tolerance and scalability of PoCL over PBFT in large-scale distributed networks.

\subsection{Cache Hit Rate}
Consider the DON layer comprising $M = 4$ oracles, each with a local cache size of $C_{m} = 4$.
During training, the exploration rate decays linearly from 0.9 to 0.05 over $T = 10,000$ time slots.
The learning rate is configured as 0.001, and the discount factor $\gamma$ is set to $0.99$.
\begin{figure*}[!t]
	\centering
	\includegraphics[width=\textwidth]{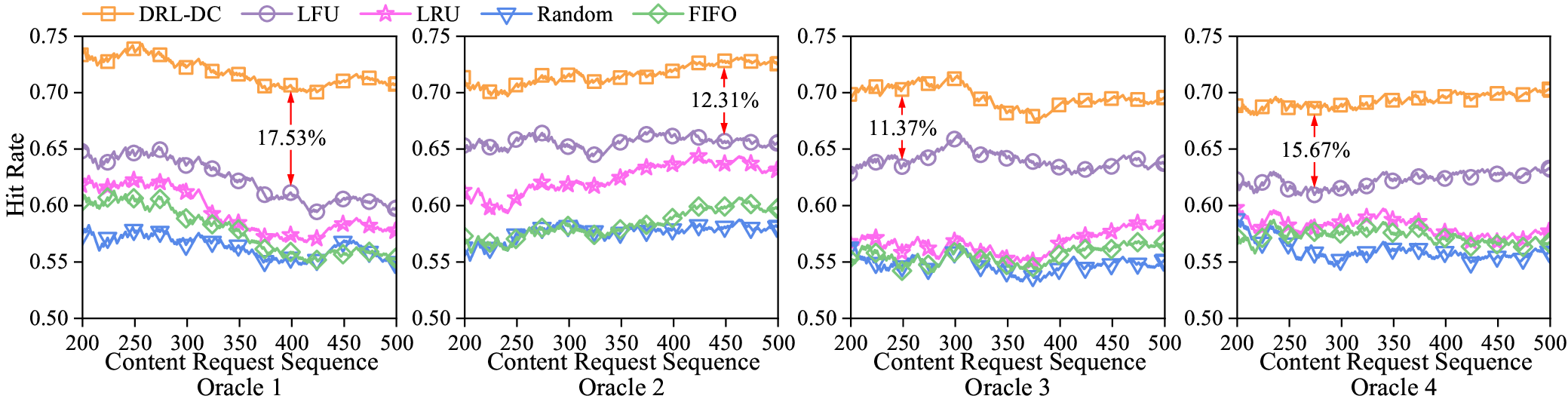}
	\caption{Comparison of caching hit rate for DRL-DC and baseline strategies under $M = 4$, Zipf $\alpha = 1.0$, and cache capacity $C_{m} = 4$.}
	\label{fig:Figure_Agent4}
\end{figure*}

Fig. \ref{fig:Figure_Agent4} shows the hit rates of DRL-DC, LFU, LRU, Random, and FIFO under different content request sequences $\boldsymbol{G}^t = \langle G_1^t, G_2^t, G_3^t, G_4^t \rangle$ with $\alpha = 1.0$ across four oracles.
The horizontal axis represents a portion of the content request sequence between the 200th and 500th requests.
First, it is observed that DRL-DC consistently reaches the highest hit rate of approximately 0.75, with a 10\% to 20\% improvement over the second-best strategy.
Second, compared with the benchmarking strategies, DRL-DC maintains the most stable performance across the content sequence with minimal fluctuations.
Finally, in oracles 2 and 4, the hit rate of DRL-DC steadily increases, while those of other strategies gradually decline.
These observations demonstrate superior adaptability and efficiency of DRL-DC in optimizing cache hit rates over traditional caching strategies.

\begin{figure}[!t]
	\centering
	\includegraphics[width=2.5in]{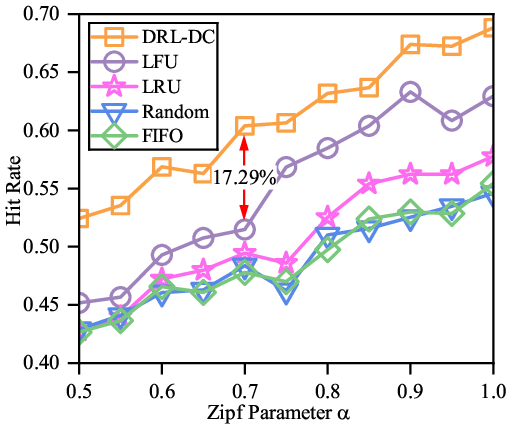}
	\caption{Comparison of caching hit rate across caching strategies with different Zipf parameters.}
	\label{fig:Comparison_Alpha}
\end{figure}

Fig. \ref{fig:Comparison_Alpha} compares the hit rates between DRL-DC, LFU, LRU, Random, and FIFO across different Zipf parameters $\alpha$.
It is observed that as $\alpha$ gradually increases, the hit rates of all strategies improve, with DRL-DC consistently achieving the highest performance.
Specifically, DRL-DC achieves a hit rate of approximately 0.55 at $\alpha = 0.5$ and approaches 0.7 at $\alpha = 1.0$, which are significantly higher than those of the baseline strategies.
These observations highlight the adaptability and superior performance of DRL-DC over traditional caching strategies under various content popularity distributions.

\subsection{Content Retrieval Latency}

We evaluate the content retrieval latency of TDC-Cache under different content popularity distributions and caching strategies.
The experiments are conducted on a DLT layer with an average transmission rate of $\bar{v} = 1$ MB/s, a content addressing time of $T_{g_{{m}, {u},k}^t} = 200$ ms, a chunk addressing time of $T_\text{chunk} = 100$ ms, and a chunk size of $C_\text{chunk} = 256$ KB.
The DON layer consists of $M=4$ oracles, each with a local cache size of $C_{m} = 4$.
The network includes $U=10$ CRs, each generating $\lvert G_{{m}, {u}}^t \rvert = 1000$ content requests per time slot.
During the PoCL consensus process, the \textbf{Prepare} and \textbf{Train} phases span $T_\text{pre} = 100$ ms and $T_\text{tra}= 100,000$ ms per time slot, respectively.

\begin{table}[!t]
	\centering
	\caption{Content Retrieval Latency (ms/KB) Under Different Content Popularity ($\alpha$ = 0.5, 1.0, and 1.5).}
    \label{tab:retrieval_latency}
	\begin{tabular}{ccccc}
	\toprule
	\multirow{2}{*}{Popularity} & \multirow{2}{*}{Strategy} & \multicolumn{3}{c}{Content Retrieval Latency (ms/KB)} \\
	\cmidrule(lr){3-5}
	 &  & 1KB--256KB & 256KB--1MB & 1MB--10MB \\
	\midrule
	\multirow{5}{*}{$\alpha = 0.5$}  & DRL-DC & \textbf{6.90} & \textbf{1.59}& \textbf{1.24}\\
                                     & LFU    & 7.13& 1.64& 1.28\\
	                                 & LRU & 7.16& 1.65& 1.29\\
	                                 & Random & 7.15& 1.65& 1.29\\
	                                 & FIFO & 7.17& 1.65& 1.29\\
	\midrule
	\multirow{5}{*}{$\alpha = 1.0$}  & DRL-DC & \textbf{6.52}& \textbf{1.52}& \textbf{1.17}\\
	                                 & LFU & 6.67& 1.55& 1.21\\
	                                 & LRU & 6.82& 1.57& 1.23\\
	                                 & Random & 6.85& 1.59& 1.24\\
	                                 & FIFO & 6.85& 1.58& 1.23\\
	\midrule
	\multirow{5}{*}{$\alpha = 1.5$}  & DRL-DC &\textbf{6.19}& \textbf{1.44}& \textbf{1.11}\\
	                                 & LFU & 6.23& 1.46& 1.13\\
	                                 & LRU & 6.32& 1.48& 1.15\\
	                                 & Random & 6.47& 1.50& 1.17\\
	                                 & FIFO & 6.50& 1.51& 1.17\\
	\midrule
	\multicolumn{2}{c}{Direct Access}& 8.19& 1.86& 1.46 \\
	\bottomrule
	\end{tabular}
\end{table}

Table \ref{tab:retrieval_latency} summarizes the content retrieval latency of TDC-Cache with DRL-DC, LFU, LRU, Random, and FIFO strategies across different content sizes and popularity levels.
From this table the proposed framework reduces retrieval latency compared with direct access, with DRL-DC consistently achieving the lowest latency across all conditions.
For example, at $\alpha = 0.5$, DRL-DC achieves 6.90 ms/KB for content sizes of 1 KB–256 KB, compared with 7.13 ms/KB for LFU, 7.16 ms/KB for LRU, 7.15 ms/KB for Random, and 7.17 ms/KB for FIFO.
For larger content sizes (e.g., above 1 MB), DRL-DC maintains the lowest latency of 1.24 ms/KB, while other strategies range from 1.28 ms/KB to 1.29 ms/KB.
As $\alpha$ increases, the performance advantage of DRL-DC becomes more pronounced.

\section{Conclusion}
\label{sec:Conclusion}
In this paper, we have proposed a novel TDC-Cache framework to address the challenges of efficient content delivery in Web3.0.
This framework establishes the DON layer and utilizes a DRL-based approach to enable decentralized cooperative caching among oracles.
Specifically, the proposed DRL-DC strategy optimizes decentralized caching decisions to improve cache hit rates and reduce content retrieval latency, while PoCL ensures the consistency of decentralized caching decisions within DON.
Extensive experiments demonstrate the superior performance of TDC-Cache over traditional caching strategies.
In particular, TDC-Cache consistently achieves higher cache hit rates and lower retrieval latency, especially under varying content popularity distributions.
The proposed framework is highly scalable and fault-tolerant, making it suitable for large-scale decentralized applications.
Future research directions include investigating incentive mechanisms to encourage honest oracle participation and addressing the challenges of dynamic network topologies.

\bibliographystyle{IEEEtran}
\bibliography{references}

\end{document}